\documentclass[a4paper]{jpconf}
\usepackage{amsmath,amssymb,amsfonts,xspace}
\usepackage{graphicx,overpic}
\usepackage{esvect,bm}
\usepackage{cite}
\newcommand{\overbar}[1]{\mkern 1.5mu\overline{\mkern-1.5mu#1\mkern-1.5mu}\mkern 1.5mu}
\newcommand{\Br}{\mathop\mathrm{Br}}
\newcommand{\RD}{\ensuremath{R_{D}}\xspace}
\newcommand{\jpsi}{\ensuremath{J\!/\!\psi}\xspace}
\newcommand{\RDst}{\ensuremath{R_{D^*}}\xspace}
\newcommand{\RDDst}{\ensuremath{R_{D^{\left(*\right)}}}\xspace}

\newcommand{\RK}{\ensuremath{R_K}\xspace}
\newcommand{\RKst}{\ensuremath{R_{K^*}}\xspace}

\newcommand{\nubar}{\ensuremath{\overbar{\nu}}\xspace}

\begin{document}
\title{Flavour anomalies: a review}

\author{Elena Graverini{\normalfont, on behalf of the ATLAS, CMS and LHCb collaborations}}

\address{Physik-Institut, University of Zurich. Winterthurerstrasse 190, 8057 Zurich}

\ead{elena.graverini@cern.ch}

\begin{abstract}
The concept of lepton flavour universality (LFU), according to which the three lepton families are equivalent except for their masses, is a cornerstone prediction of the Standard Model (SM). LFU can be violated in models beyond the SM by new physics particles that couple preferentially to certain generations of leptons. In the last few years, hints of LFU violation have been observed in both tree-level $b\to c\ell\nu$ and loop-level $b\to s\ell\ell$ transitions. These measurements, combined with the tensions observed in angular observables and branching fractions of rare semileptonic $b$ decays, point to a coherent pattern of anomalies that could soon turn into the first observation of physics beyond the SM. These proceedings review the anomalies seen by the LHC experiments and the $B$ factories, and give an outlook for the near future.
\end{abstract}

\section{Introduction}
Quarks and leptons, the fundamental fermions of the SM, exist in three generations, each comprised of two
members.
The property that distinguishes the fundamental fermions from one another is called flavour.
The universality of lepton flavour (LFU) is one of the most interesting consequences of the Standard Model (SM).
LFU is an accidental symmetry, broken only by Yukawa interactions, and states that the electroweak gauge bosons couple with equal strength to the three families of leptons.
This property is well established in decays of light mesons, \textit{e.g.} $K\to\ell\nu$ decays~\cite{Lazzeroni:2012cx}.
A violation of LFU would clearly indicate that new particles participate in quark flavour changing processes, modifying their dynamics.

In the SM, transitions between different quark flavours can only be mediated by the charged weak bosons $W^\pm$.
As a consequence, flavour-changing neutral current (FCNC) transitions between same-charge quarks are not directly
mediated by the neutral weak boson $Z^0$, but rather occur through much rarer loop processes involving virtual
$W^\pm$ and additional virtual quarks, in penguin- and box-like Feynman diagrams.
The SM predicts the dynamics of decays governed by FCNC transitions with very high precision.
New particles can either participate in the loops, or generate additional tree-level diagrams.
The amplitudes of suppressed decays governed by $b\to s\ell\ell$ transitions are ideal laboratories to look for New Physics (NP), as effects beyond the SM can be sizeable with respect to the competing SM processes.
An intriguing set of anomalies emerged in recent years from the study of such amplitudes~\cite{Aaij:2014pli,Aaij:2015xza,Aaij:2015esa,Aaij:2016flj,Aaij:2014ora,Wei:2009zv,Lees:2012tva,Aaij:2017vbb,Aaij:2015oid,Wehle:2016yoi,Aaboud:2018krd,Sirunyan:2017dhj}.
In addition, tree-level decays of beauty mesons to final states with a $\tau$ lepton have been studied at BaBar~\cite{Lees:2012xj,Lees:2013uzd}, Belle~\cite{Huschle:2015rga,Abdesselam:2016cgx,Hirose:2016wfn} and LHCb~\cite{Aaij:2015yra,Aaij:2017uff}. In all cases, hints of a deviation from LFU were reported.
A coherent picture emerges, with many NP models predicting particles with enhanced couplings to the second and third generation of quarks and leptons, see \textit{e.g.} \cite{Buttazzo:2017ixm,Wei:2017ago,Bauer:2015knc,Greljo:2015mma,Barbieri:2017tuq,Bordone:2017bld}.

In this proceedings, recent results in $b$-hadron decays are reviewed, with a focus on tests of LFU and angular analyses performed at the Large Hadron Collider (LHC).

\section{Lepton universality in charged current transitions}
\begin{figure}[tp]
    \centering
    \includegraphics[width=.5\textwidth]{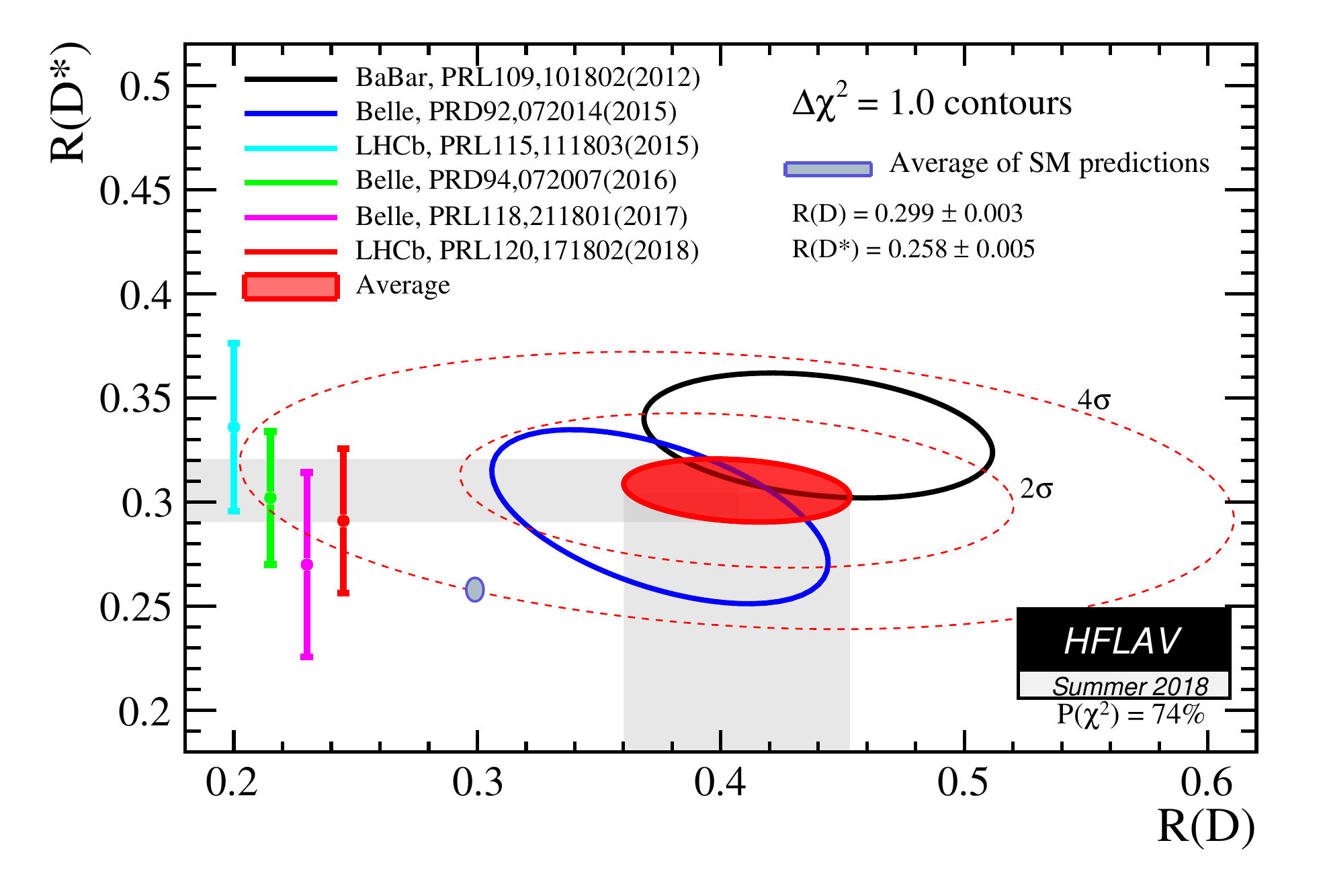}
    \caption{Averages of the \RDDst ratios as computed in summer 2018~\cite{hflav}.}\label{img:rlcs:hflav}
\end{figure}

The rates of $b$-meson decays to $\tau$ and $\mu$ leptons are expected to differ because of the substantial $\mu$-$\tau$ mass difference.
$B\to D^{\left(*\right)}\ell\nubar_\ell$ decays have been studied at BaBar~\cite{Lees:2012xj,Lees:2013uzd}, Belle~\cite{Huschle:2015rga,Abdesselam:2016cgx,Hirose:2016wfn} and LHCb~\cite{Aaij:2015yra,Aaij:2017uff}. In all cases, the measured observables
\begin{align}
    \RDDst \equiv \frac{\Br\big(B\to D^{\left(*\right)}\tau\nubar_\tau\big)}{\Br\big(B\to D^{\left(*\right)}\ell\nubar_\ell\big)},
    \quad \text{ with } \ell=\mu, e
\end{align}
consistently exceed SM expectations.
\figurename~\ref{img:rlcs:hflav} shows a combination of these experimental results, and compares them with the most recently calculated SM predictions~\cite{Bigi:2017jbd,Jaiswal:2017rve}.
The following averages are obtained~\cite{hflav}:
\begin{align}
    &\RD = 0.407 \pm 0.039 \text{ (stat)} \pm 0.024 \text{ (syst)}
    &R_{D,\,SM} = 0.299 \pm 0.003\\
    &\RDst = 0.306 \pm 0.013 \text{ (stat)} \pm 0.007 \text{ (syst)}
    &R_{D^*,\,SM} = 0.258 \pm 0.005
\end{align}
The experimental values of \RD and \RDst exceed the SM expectactions by $2.3$ and $3.0$ standard deviations ($\sigma$), respectively, for a resulting combined tension with the SM of about $3.8\sigma$.

The fact that this discrepancy has been observed both at $B$-factory experiments and at the LHC corroborates its significance. 
The Belle and BaBar experiments measured the semitauonic decay rate relative to the total $e,\,\mu$ rate.
Due to the difficult reconstruction of electrons, LHCb only used the semimuonic channel as normalization.
Whereas the $B$ meson kinematics is completely known at $B$-factories, at the LHC the momenta of the colliding partons are unknown.
Together with the presence of invisible neutrinos in the final state, this makes the measurement of semileptonic $B$ decays very challenging at a hadron collider.
Owing to the excellent resolution of its vertex detector, LHCb manages to reconstruct the flight direction and momentum of the decaying $B$ meson with a resolution of about 18\%~\cite{Aaij:2015yra}.
The longitudinal boost of the $B$ meson is assumed to be well approximated by that of the reconstructed meson-lepton pair.

\begin{figure}
\begin{overpic}[width=.4\textwidth]{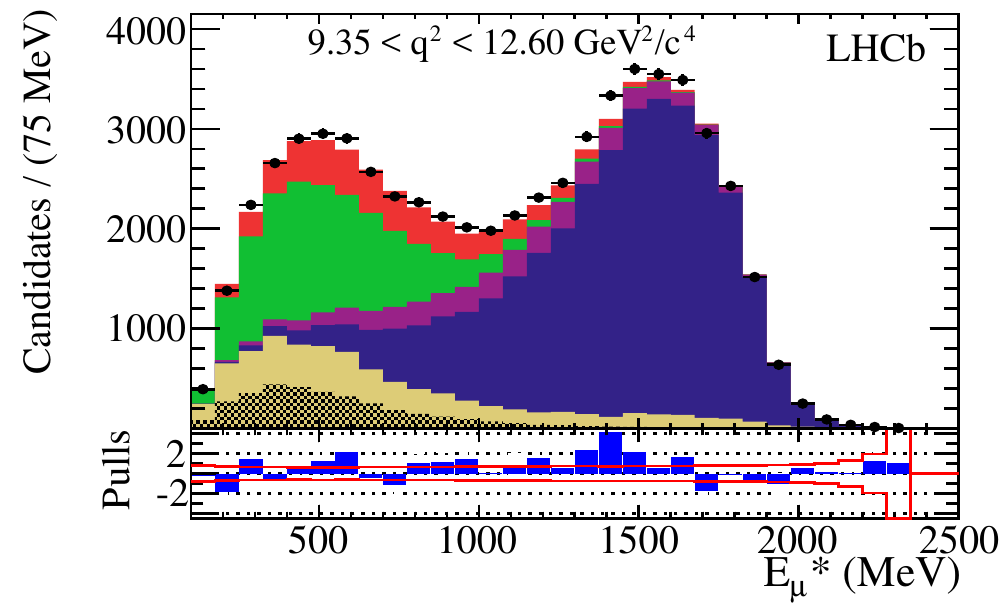}\end{overpic}\hspace*{-2mm}
\begin{overpic}[width=.4\textwidth]{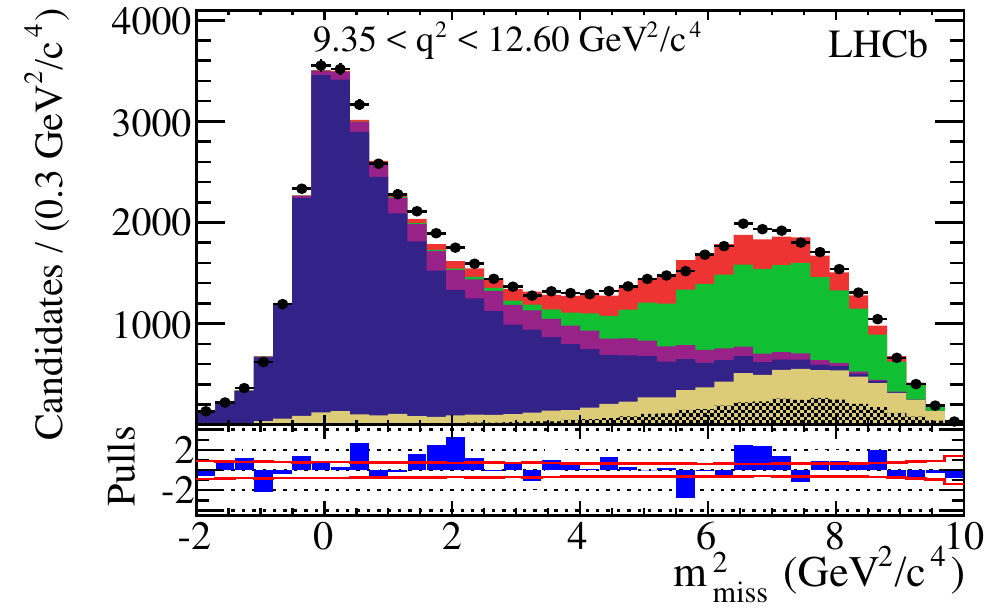}\put(101,15){\includegraphics[width=.2\textwidth]{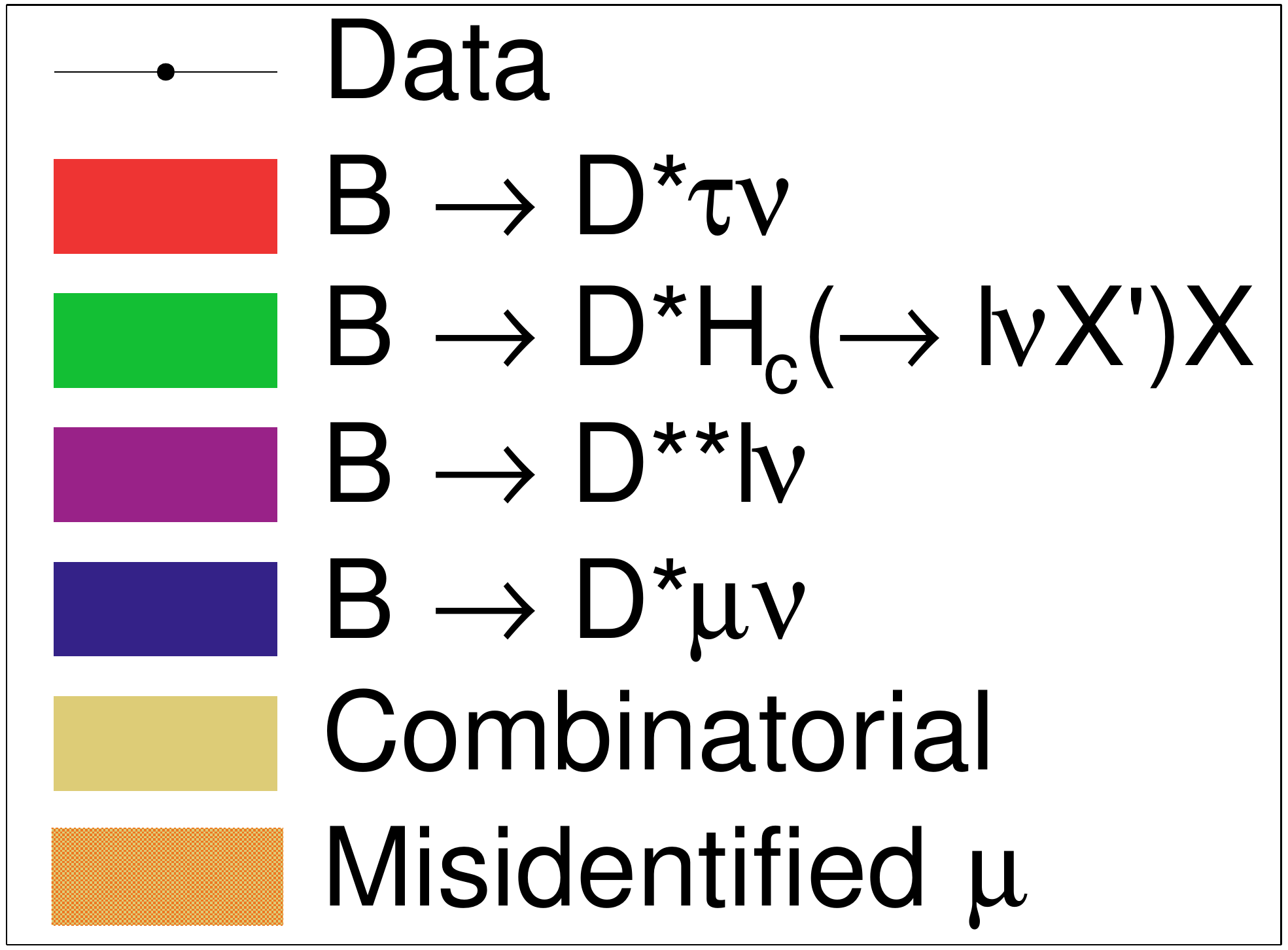}}\end{overpic}\\[+7pt]
\begin{overpic}[width=.329\textwidth]{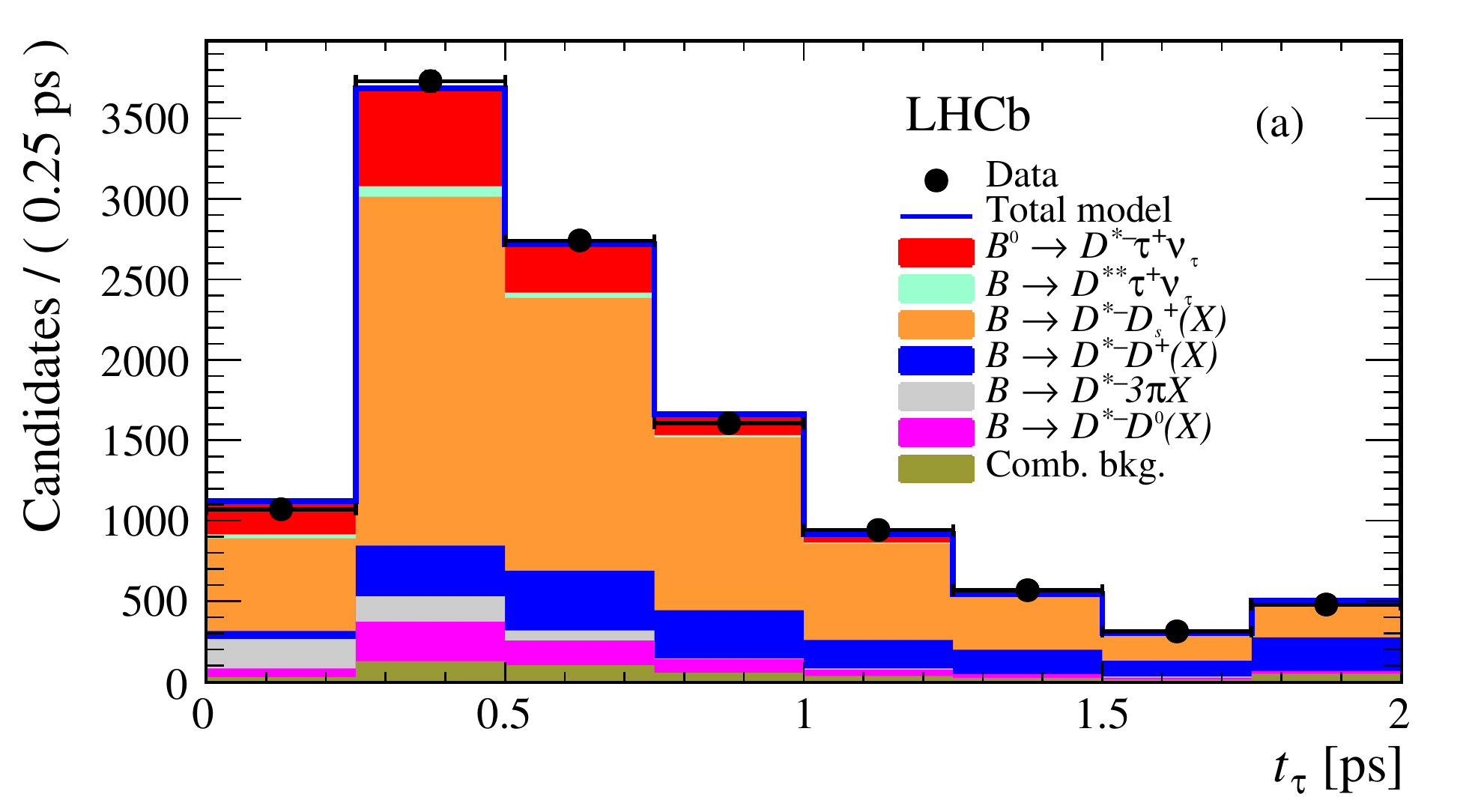}\end{overpic}
\begin{overpic}[width=.329\textwidth]{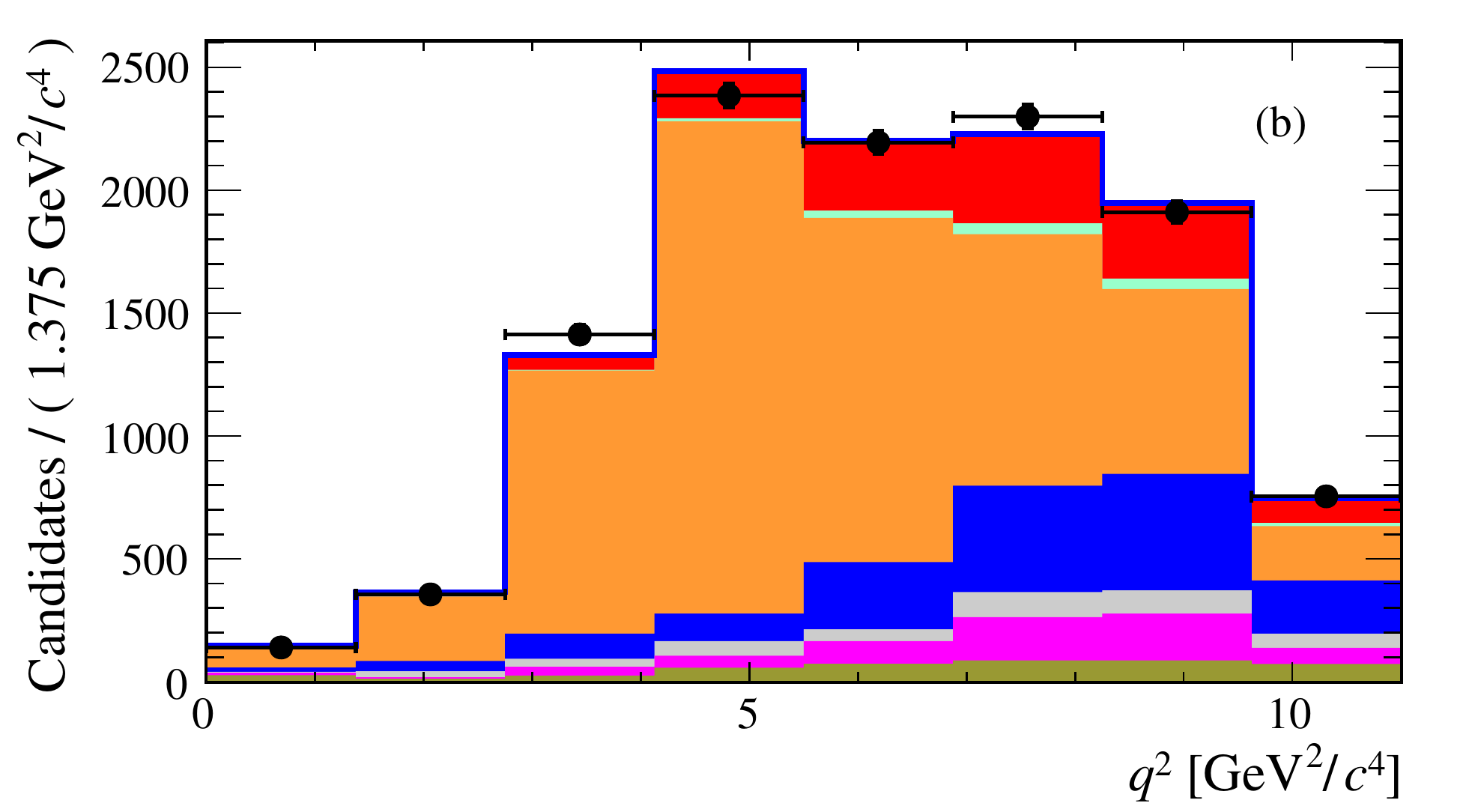}\end{overpic}
\begin{overpic}[width=.329\textwidth]{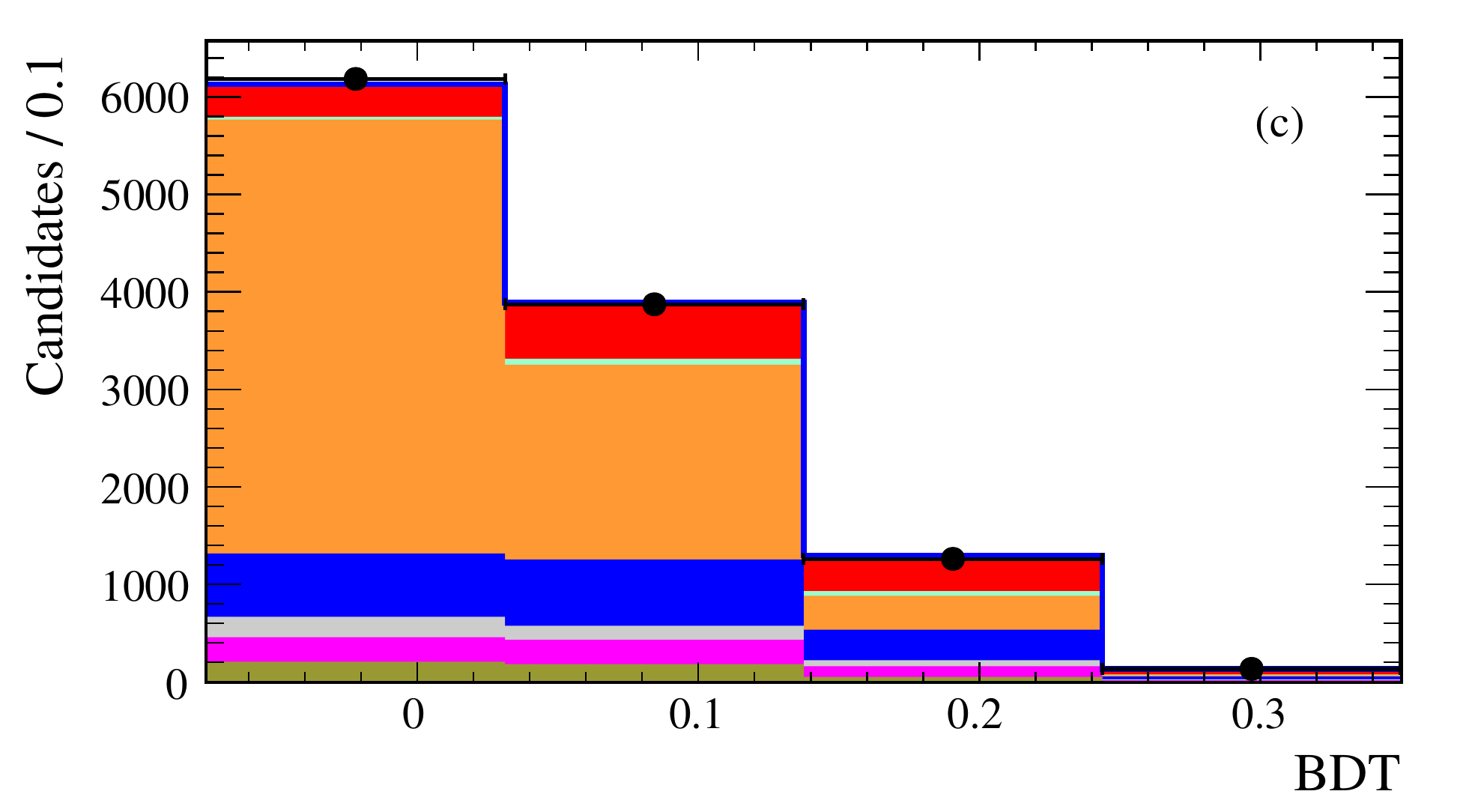}\end{overpic}
\caption{$E_\mu^*$ and $m^2_{miss}$ spectra for $9.35\leq q^2\leq 12.60$~GeV$^2/c^4$ in the LHCb measurement of \RDst{} using semileptonic $\tau$ decays (top panels)~\cite{Aaij:2015yra}. Distributions of the $\tau$ lifetime, $q^2$ and BDT output fitted in the LHCb measurement of \RDst{} using hadronic $\tau$ decays (bottom panels)~\cite{Aaij:2017uff}.}\label{fig:Rdst}
\end{figure}

LHCb performed a first measurement of \RDst{} using leptonic $\tau^+\to\mu^+\nu_\mu\nubar_\tau$ decays (charge-conjugate modes are implied hereinafter).
Many systematic uncertainties cancel by measuring decays with the same reconstructed particles ($D^{*-}$ and $\mu^+$).
A three-dimensional template fit is used to separate the two final states $D^{*-}\mu^+\nu_\mu\nu_\tau\nubar_\tau$ and $D^{*-}\mu^+\nu_\mu$.
Variables sensitive to the number of neutrinos in the final states are used to calculate the relative yields of semitauonic and semimuonic decays: the missing mass squared $m^2_{miss}$, the squared $B\to D^*$ recoil $q^2$ and the rest frame muon energy $E_\mu^*$.
Simulation validated against data is used to estimate the shape of physics backgrounds such as $B^0\to D^{*-} D^+$, whereas all other backgrounds are evaluated using data-driven templates.
The top panels of \figurename~\ref{fig:Rdst} show the $E_\mu^*$ and $m^2_{miss}$ spectra in the highest $q^2$ bin, where the semitauonic contribution is largest.
The LFU observable was measured to be $\RDst{}=0.336 \pm 0.027 \text{ (stat)} \pm 0.030 \text{ (syst)}$, compatible with the SM within $2.1\sigma$~\cite{Aaij:2015yra}.

A subsequent measurement from LHCb used hadronic $\tau^-\to\pi^+\pi^-\pi^-(\pi^0)\nu_\tau$ decays. The branching fraction $\Br\left(B^0\to D^{*-}\tau^+\nu_\tau\right)$ was measured relative to that of $B^0\to D^{*-}{(3\pi)}^+$, which is well known~\cite{Tanabashi:2018}.
As in the previous case, this strategy allows to suppress many reconstruction-related systematic uncertainties.
The relatively long $\tau$ lifetime is exploited to suppress background from prompt pions, \textit{i.e.} $B^0\to D^{*-}\,3\pi X$.
A multivariate approach is adopted to suppress background from decays with an additional charmed meson, using a Boosted Decision Tree (BDT)~\cite{Breiman:1984jka}.
The signal yield is determined with a three-dimensional fit to the $\tau$ lifetime, the output of the BDT and $q^2$.
The bottom panels of \figurename~\ref{fig:Rdst} show the result of the fit, highlighting the various contributions to the measured signal yield.
The known branching fractions $\Br\left(B^0\to D^{*-}{(3\pi)}^+\right)$ and $\Br\left(B^0\to D^{*-}\mu^+\nu_\mu\right)$ are used to calculate \RDst{}, which is measured to be $\RDst = 0.291 \pm 0.019 \text{ (stat)} \pm 0.026 \text{ (syst)} \pm 0.013 \text{ (ext)}$, where the last uncertainty is due to the external input brancing fractions~\cite{Aaij:2017uff}.
This result is compatible with the SM within $1\sigma$, and lies slightly above the prediction, as in the case of~\cite{Aaij:2015yra}.

Lepton flavour universality can also be tested in $B_c$ decays.
While the $B$-factories operate on the $\Upsilon(4S)$ resonance for a majority of their data taking,
measurements using other $B_q$ species are possible at the LHC.
LHCb performed a measurement of the ratio
\begin{align}
R_{\jpsi} \equiv \frac{\Br\big(B_c^+\to \jpsi\,\tau^+\nu_\tau\big)}{\Br\big(B_c^+\to \jpsi\,\mu^+\nu_\mu\big)}
\end{align}
with the $\tau^+$ decaying leptonically to $\mu^+\nu_\mu\nubar_\tau$.
The theoretical uncertainties on the form factors governing the $B_c\to \jpsi$ transition result in large uncertainties on the SM predictions for $R_{\jpsi}$, with central values lying in the $[0.25, 0.28]$ range~\cite{Anisimov:1998uk,Kiselev:2002vz,Ivanov:2006ni,Hernandez:2006gt}. LHCb performed a three-dimensional fit to $m^2_{miss}$, the $B_c$ lifetime, and a categorical variable $Z$ representing eight bins in $(q^2,E_\mu^*)$, finding $R_{\jpsi}=0.71 \pm 0.17 \text{ (stat)} \pm 0.18 \text{ (syst)}$, again exceeding predictions. This result is compatible with the SM within about $2\sigma$~\cite{Aaij:2017tyk}.
Further tree-level LFU tests are ongoing at LHCb, including $R_{D^+}$ and the baryonic observables $R_{\Lambda_c^{(*)}}$.

\section{Flavour anomalies in rare $\bm{b}$ decays}
Rare decays of heavy-flavoured hadrons can be described by effective Hamiltonians that encode SM and possible NP contributions in the Wilson coefficients weighting the operators participating in the process. In this framework, called Operator Product Expansion (OPE)~\cite{Wilson1972}, a model-independent analysis of effects beyond the SM is possible.
In particular, $b\to s\ell\ell$ transitions are described by the effective Hamiltonian
\begin{align}
\mathcal{H}_{ef\!f} = -\frac{4G_F}{\sqrt{2}}V_{tb}V_{ts}^*\sum_i\left(\mathcal{C}_i\mathcal{O}_i + \mathcal{C}_i^\prime\mathcal{O}_i^\prime\right),
\end{align}
where $G_F$ is the Fermi constant, $V_{ij}$ are elements of the CKM matrix~\cite{Cabibbo:1963yz,Kobayashi:1973fv},
$\mathcal{O}_i^{(\prime)}$ are local operators encoding left(right)-handed long distance contributions, and $\mathcal{C}_i^{(\prime)}$ are the corresponding Wilson coefficients.

\begin{figure}
\centering
\raisebox{-0.5\height}{\begin{overpic}[width=.325\textwidth]{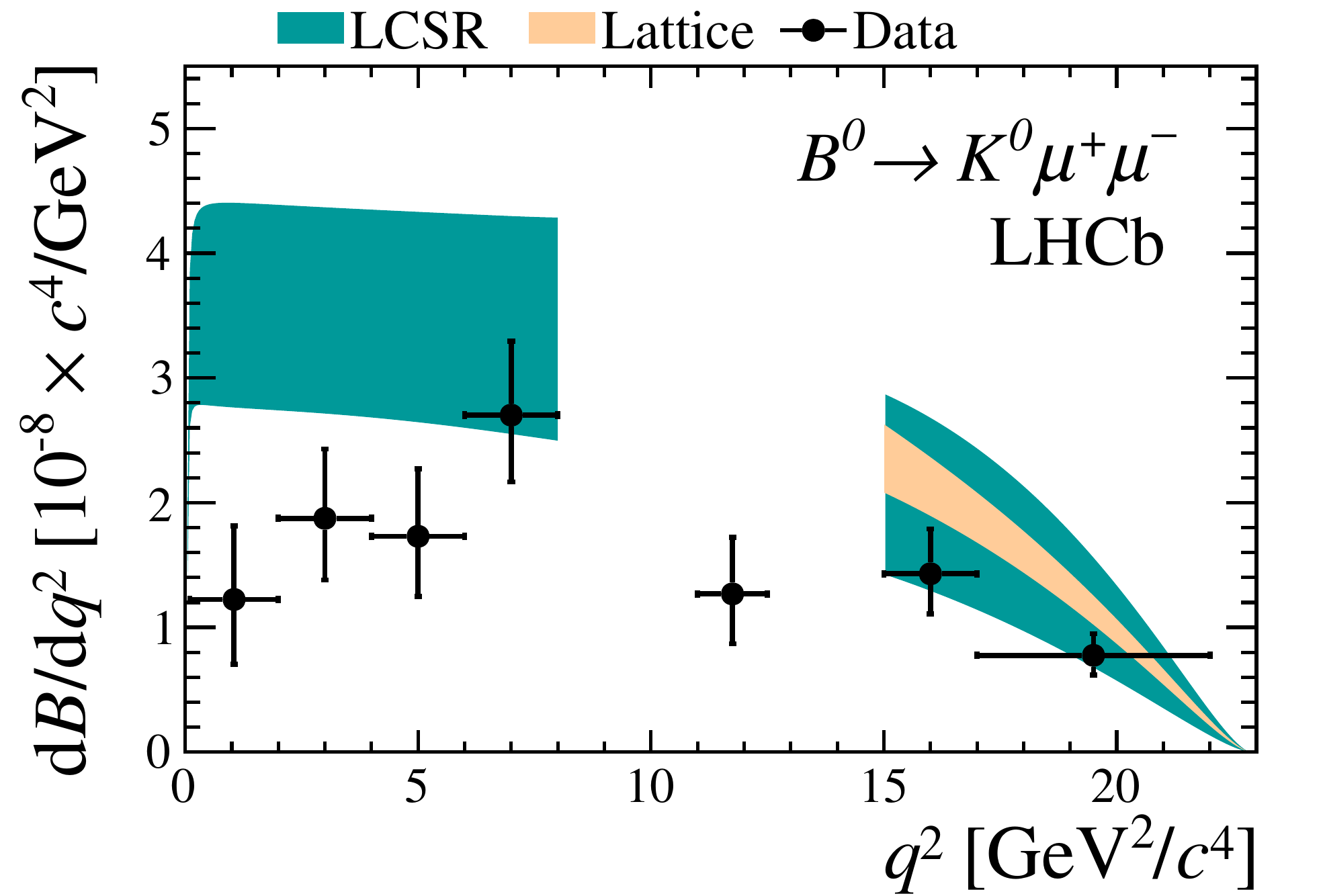}\end{overpic}}
\raisebox{-0.5\height}{\begin{overpic}[width=.325\textwidth]{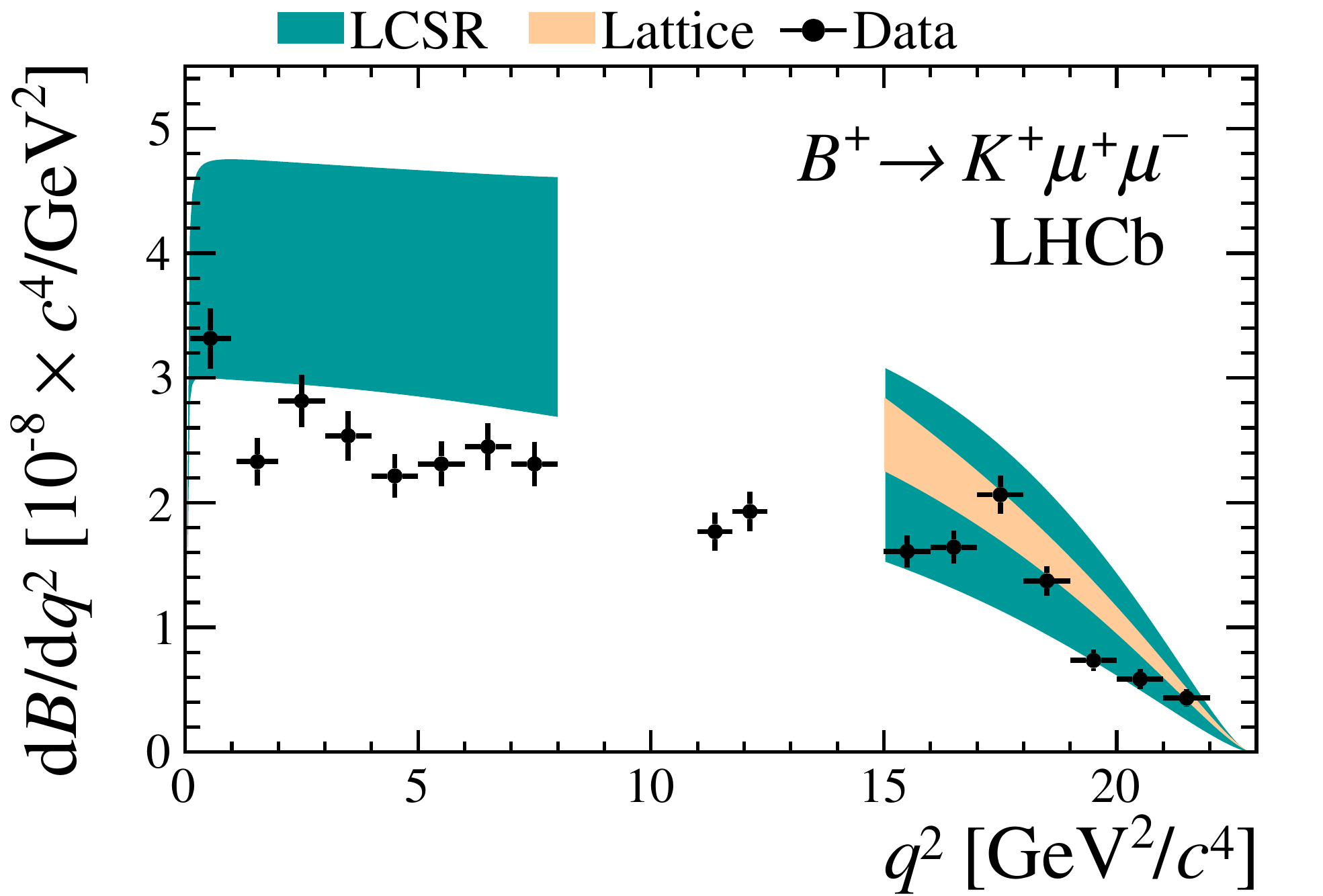}\end{overpic}}
\raisebox{-0.5\height}{\begin{overpic}[width=.325\textwidth]{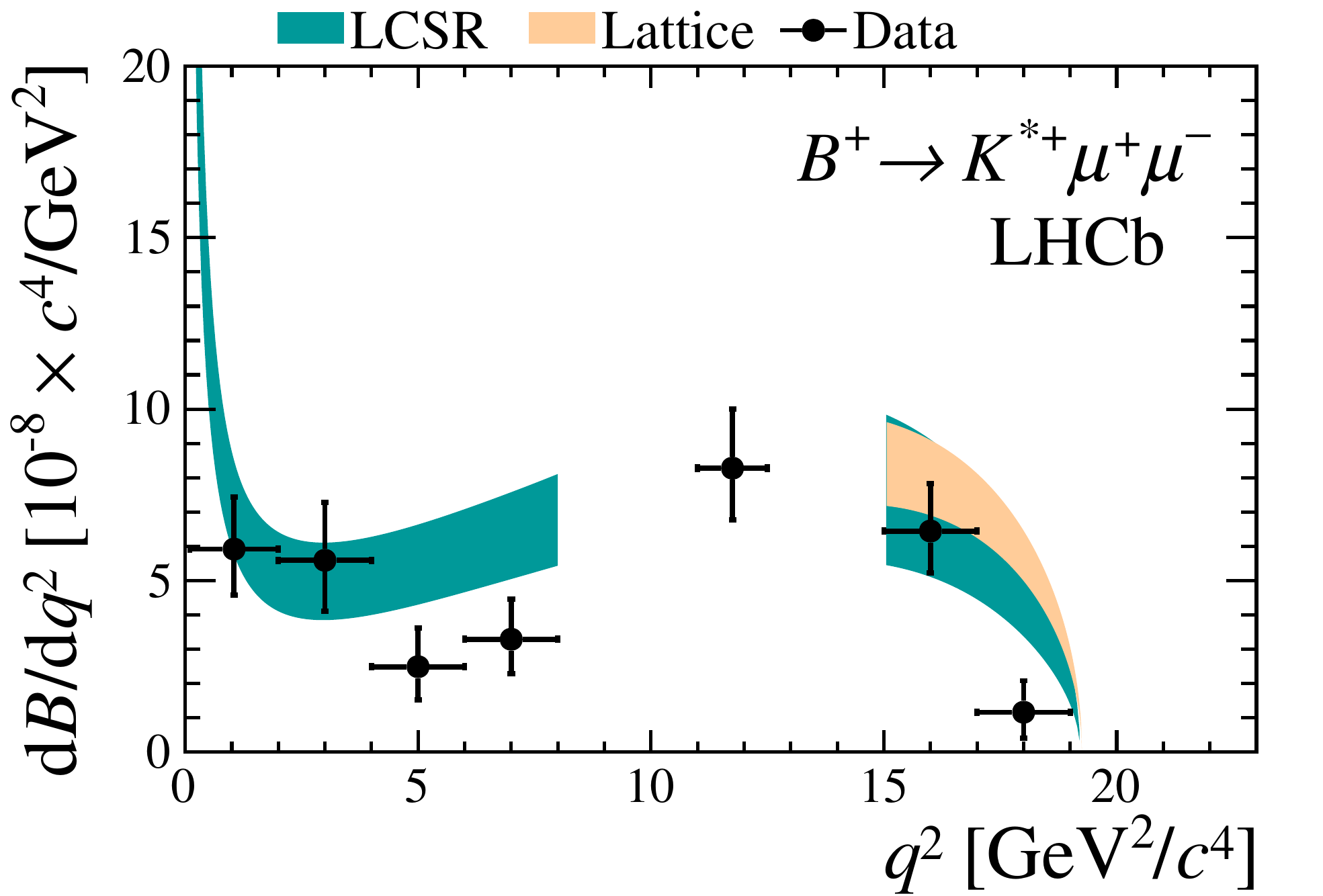}\end{overpic}}\\
\raisebox{-0.5\height}{\begin{overpic}[width=.325\textwidth]{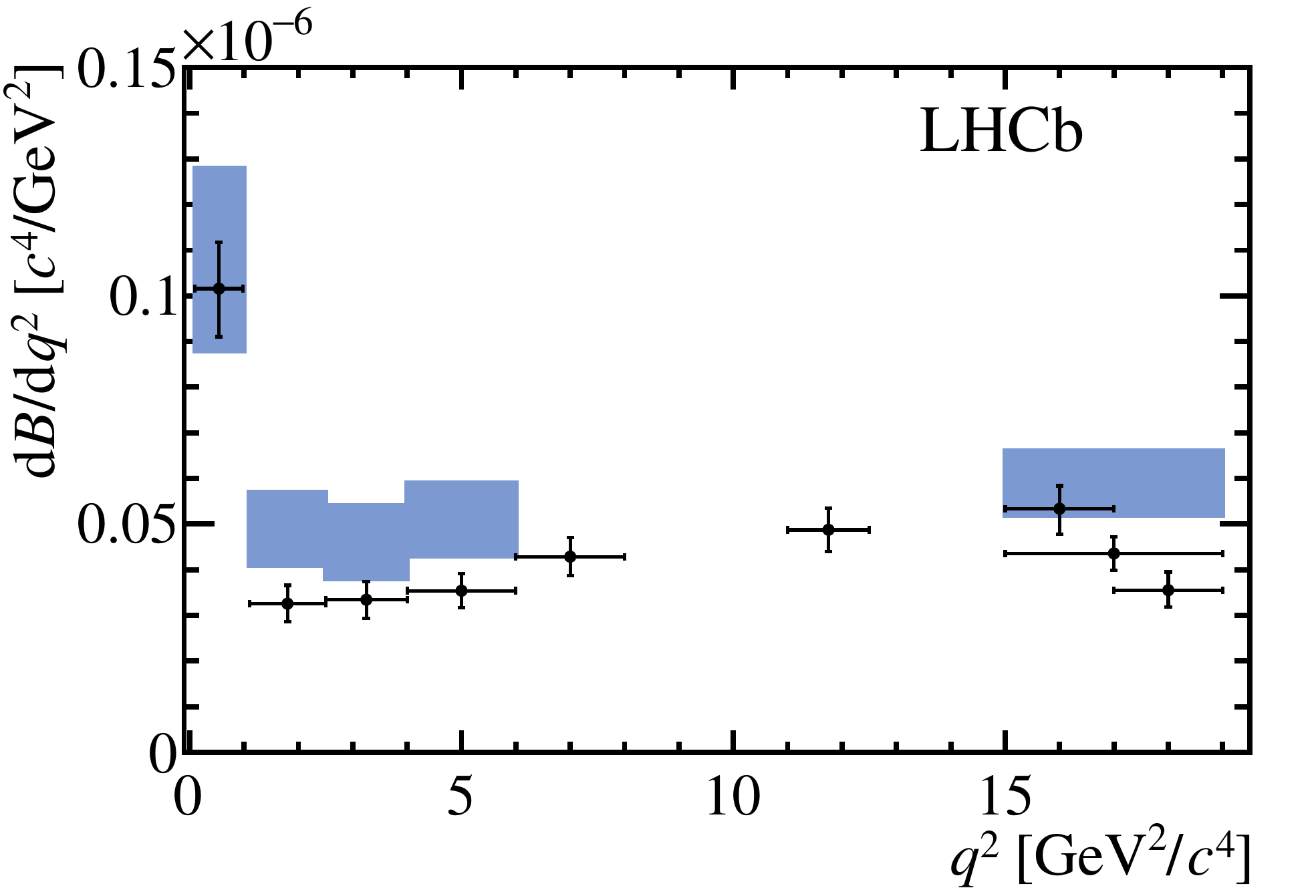}\put(54,50){\tiny$\bm{B^0\to K^{*0}\mu\mu}$}\end{overpic}}
\raisebox{-0.5\height}{\begin{overpic}[width=.325\textwidth]{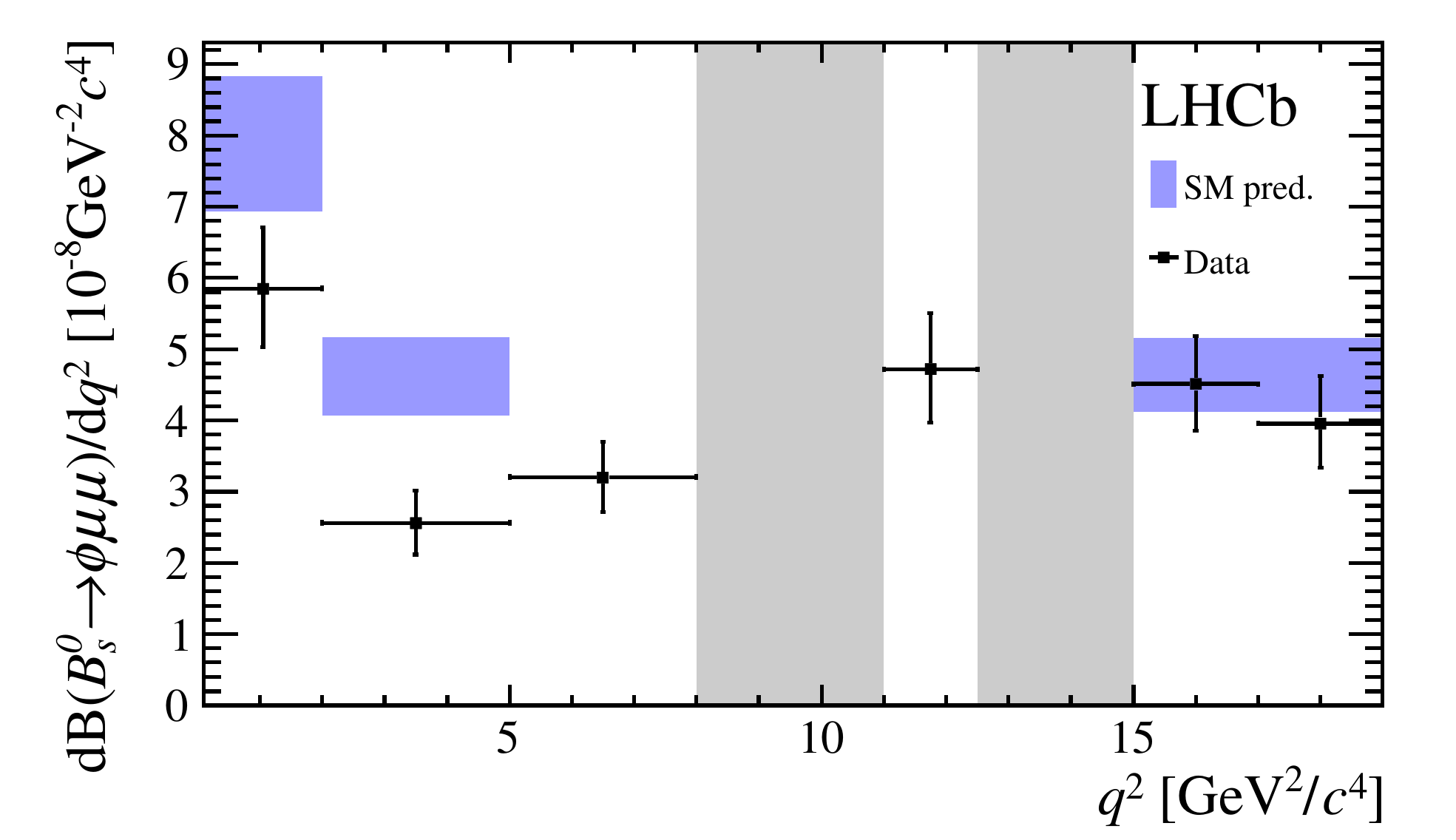}\put(20,58){\tiny$\bm{B^0_s\to \phi\mu\mu}$}\end{overpic}}
\raisebox{-0.5\height}{\begin{overpic}[width=.325\textwidth]{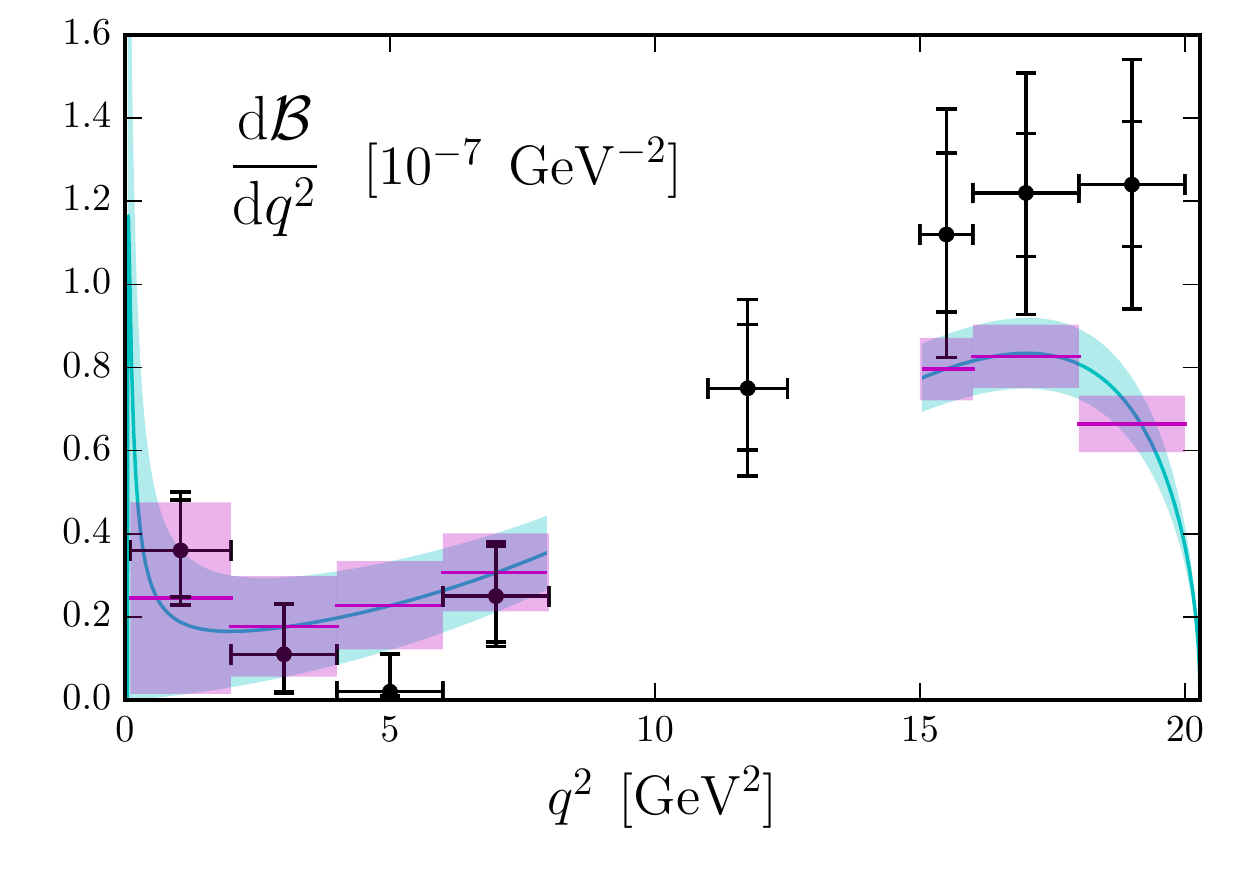}\put(55,20){\tiny$\bm{\Lambda_b^0\to \Lambda^{0}\mu\mu}$}\end{overpic}}
\caption{Differential branching fraction for various $b\to s\mu\mu$ transitions, superimposed to SM predictions~\cite{Aaij:2014pli,Aaij:2015xza,Aaij:2015esa,Aaij:2016flj,Detmold:2016pkz}.}\label{fig:bfs}
\end{figure}
Various discrepancies with the SM predictions have been detected in decays dominated by the effective vector and axial-vector couplings $\mathcal{C}_9$ and $\mathcal{C}_{10}$.
Between 2014 and 2017, branching fractions of decays such as $B^0\to K^{0}\mu^+\mu^-$, $B^0\to K^{*0}\mu^+\mu^-$, $B^+\to K^{*+}\mu^+\mu^-$, $B^0_s\to\phi\mu^+\mu^-$, $\Lambda_b^0\to \Lambda^{0}\mu^+\mu^-$, all proceeding through a $b\to s\mu\mu$ transition, have been measured at LHCb~\cite{Aaij:2014pli,Aaij:2015xza,Aaij:2015esa,Aaij:2016flj}.
For all of these channels, interenstingly, the SM expectations exceed the measured value, as visible in \figurename~\ref{fig:bfs}.
The statistical significance of these anomalies is such that a SM explanation is possible.
However, many other small discrepancies -- detailed below -- have been registered over the years, resulting altogether in a significant tension with the SM.

\subsection{Tests of LFU with $b\to s\ell\ell$ decays}
\begin{figure}
\centering
\raisebox{-.5\height}{\includegraphics[width=.49\textwidth]{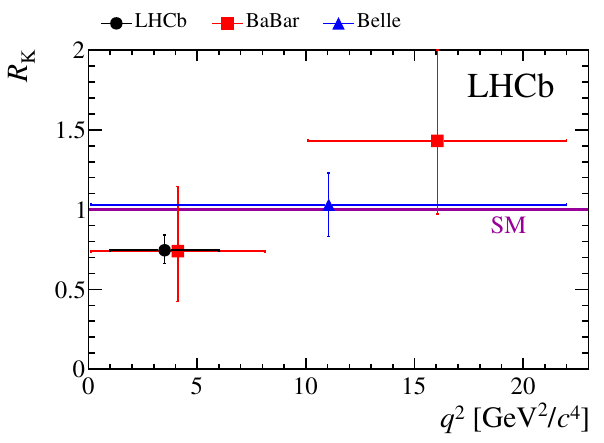}}
\raisebox{-.5\height}{\includegraphics[width=.49\textwidth]{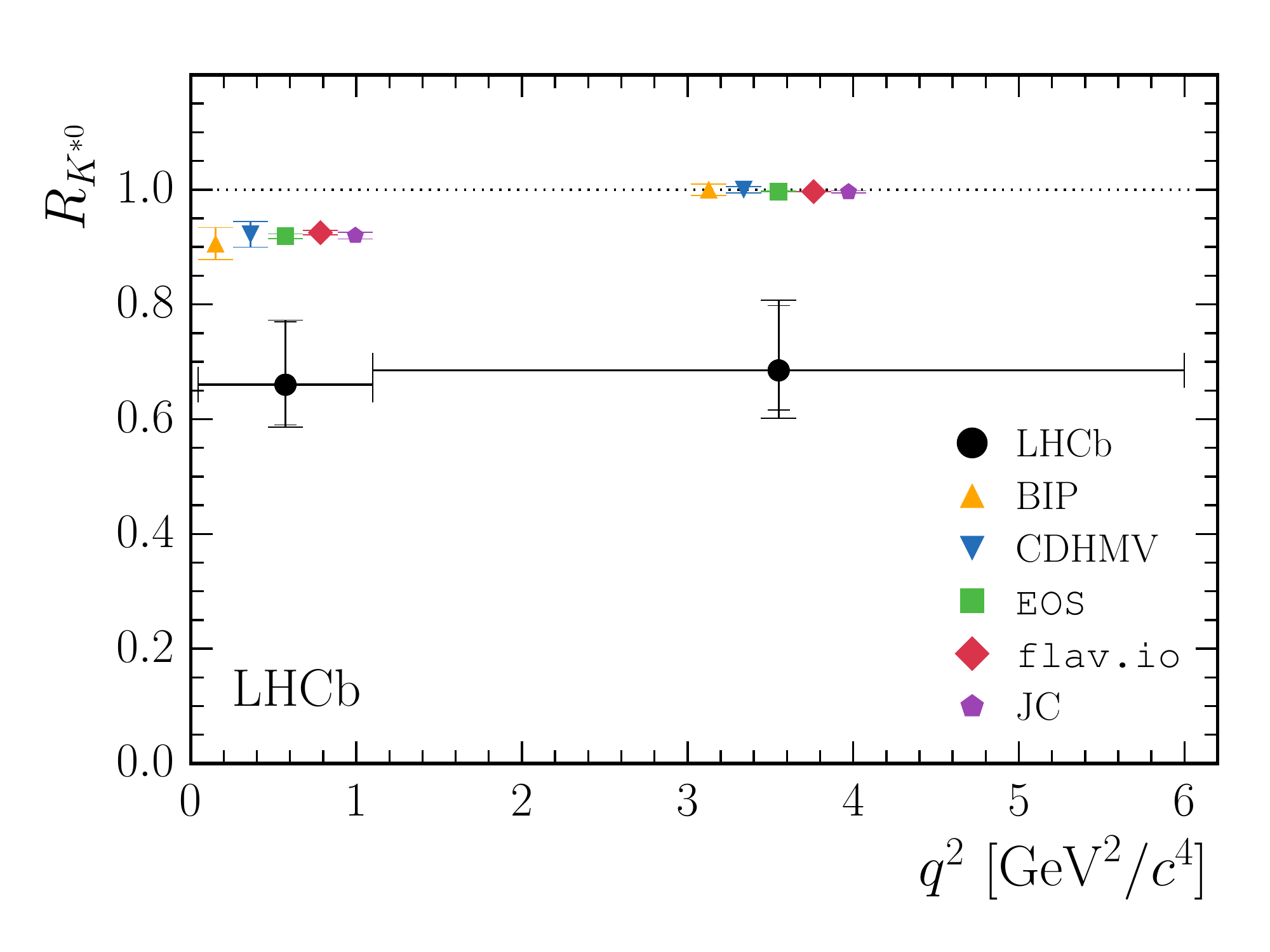}}
\caption{LHCb~\cite{Aaij:2014ora}, Belle~\cite{Wei:2009zv} and BaBar~\cite{Lees:2012tva} measurements
of \RK{} (left) and LHCb measurement of \RKst{}~\cite{Aaij:2017vbb} (right),
superimposed to SM predictions~\cite{Bordone2016,Serra:2016ivr,Altmannshofer:2017fio,
Jager:2014rwa,Capdevila:2016ivx}. Previous $\RKst{}$ measurements from Belle and BaBar can be found in~\cite{Wei:2009zv,Lees:2012tva}.}\label{fig:RKKst}
\end{figure}
Uncertainties in the hadronic form factors, and other hadronic uncertainties, cancel to a very large extent in the SM predictions for the LFU ratios
\begin{align}
R_{K^{(*)}} \equiv \frac{\Br\left(B\to K^{(*)}\mu^+\mu^-\right)}{\Br\left(B\to K^{(*)}e^+e^-\right)},
\end{align}
provided the momentum transfer $q^2$ is sufficiently larger than the dilepton mass~\cite{Bordone2016,Serra:2016ivr,Altmannshofer:2017fio,
Jager:2014rwa,Capdevila:2016ivx}. These observables are predicted to be unity with uncertainties below 1\%~\cite{Bordone2016}.
The LHCb experiment has provided experimental measurements of these quantities, laying out a common strategy for LFU tests with rare decays.
The $R_{X}$ observables are defined as ratios of efficiency corrected yields limited to certain $q^2$ ranges, chosen in order to exclude the $\jpsi$ and $\Psi(2S)$ resonances, which are then used as control channels.
Electron and muon channel yields are measured relative to the corresponding, much more abundant resonant modes $B\to X \jpsi$, where $X$ is the strange meson under study and the $\jpsi$ meson decays to either a $\mu\mu$ or $ee$ pair.
This way, thanks to the topological similarity between the nonresonant and resonant modes, the systematic uncertainties related to the differences in the reconstruction of electron and muon tracks largely cancel.

In order to test the validity of the analysis procedure, the efficiency corrected resonant yields are compared, and the important cross-check observable $r_{\jpsi} \equiv \Br\left(B\to X \jpsi\left(\to\mu\mu\right)\right) / \Br\left(B\to X \jpsi\left(\to ee\right)\right)$, expected to be unity, is measured.
This way, the electron and muon reconstruction efficiencies, as well as the efficiency of the offline selection, are validated.
The electron mode is much more challenging from an experimental point of view, and the low reconstruction efficiency for dielectron final states represents the dominant factor in the statistical uncertainty associated to the LHCb measurements.

The ratio \RK was measured with $B^+\to K^+\ell^+\ell^-$ decays in the $1.1< q^2< 6.0$~GeV$^2/c^4$ range, finding 
$\RK = 0.745 \,^{+0.090}_{-0.074} \text{ (stat)} \pm 0.036 \text{ (syst)} $, 
about $2.6\sigma$ below the SM prediction~\cite{Aaij:2014ora}.
The ratio $\RKst{}$ was later measured with $B^0\to K^{*0}\ell^+\ell^-$ decays in two disjoint $q^2$ bins, finding
\begin{align}
&\RKst{} = 0.66 \,^{+0.11}_{-0.07} \text{ (stat)} \pm 0.03 \text{ (syst)} \quad\text{ for } 0.045< q^2< 1.1~\text{GeV}^2/c^4\\
&\RKst{} = 0.69 \,^{+0.11}_{-0.07} \text{ (stat)} \pm 0.05 \text{ (syst)} \quad\text{ for } 1.1< q^2< 6.0~\text{GeV}^2/c^4
\end{align}
with a SM compatibility at the $2.2$-$2.5\sigma$ level~\cite{Aaij:2017vbb}. At the same time, the control ratio $r_{\jpsi}$ was found compatible with unity within $1\sigma$, with $r_{\jpsi}=1.043\pm 0.006\text{ (stat)}\pm 0.045\text{ (syst)}$~\cite{Aaij:2017vbb}.
The main systematic uncertainties for both ratios arise from double-misidentification of $\jpsi$ decay products, from bremsstrahlung losses affecting the $B$ mass shape in the electron channel, and from the determination of the trigger and selection efficiencies.
Some of these uncertainties also depend on the size of the simulated samples used to assess the efficiencies, and are expected to shrink if more events are simulated.
The \RK{} and \RKst{} measurements from LHCb are shown in the left- and right-hand panel of \figurename~\ref{fig:RKKst}, respectively, where they are compared to the SM predictions and to the measurements performed by the Belle~\cite{Wei:2009zv} and BaBar~\cite{Lees:2012tva} experiments. More $R_{X}$ measurements are foreseen at LHCb, using \textit{e.g.} $B^+\to K^+\pi^+\pi^-\ell^+\ell^-$, $\Lambda_b^0\to\Lambda^{\!*0}\ell^+\ell^-$ and $B_s^0\to\phi\ell^+\ell^-$ decays.

\subsection{$B^0\to K^{*0}\mu^+\mu^-$ angular analysis}
The source of the anomalies in $b\to s\ell\ell$ branching fractions and LFU observables is unclear.
If new particles were partecipating in these decays, they would be expected to modify their rates, and also the angular distribution of the decay products. The latter is in fact induced by the scalar, vector or axial-vector nature of the decay mediator(s).
For this reason, LHCb performed an analysis of the angular distribution of the particles produced in $B^0\to K^{*0}\mu^+\mu^-$ decays, using data from the LHC Run 1~\cite{Aaij:2015oid}.

\begin{figure}
\centering
\raisebox{-.5\height}{\includegraphics[width=.59\textwidth]{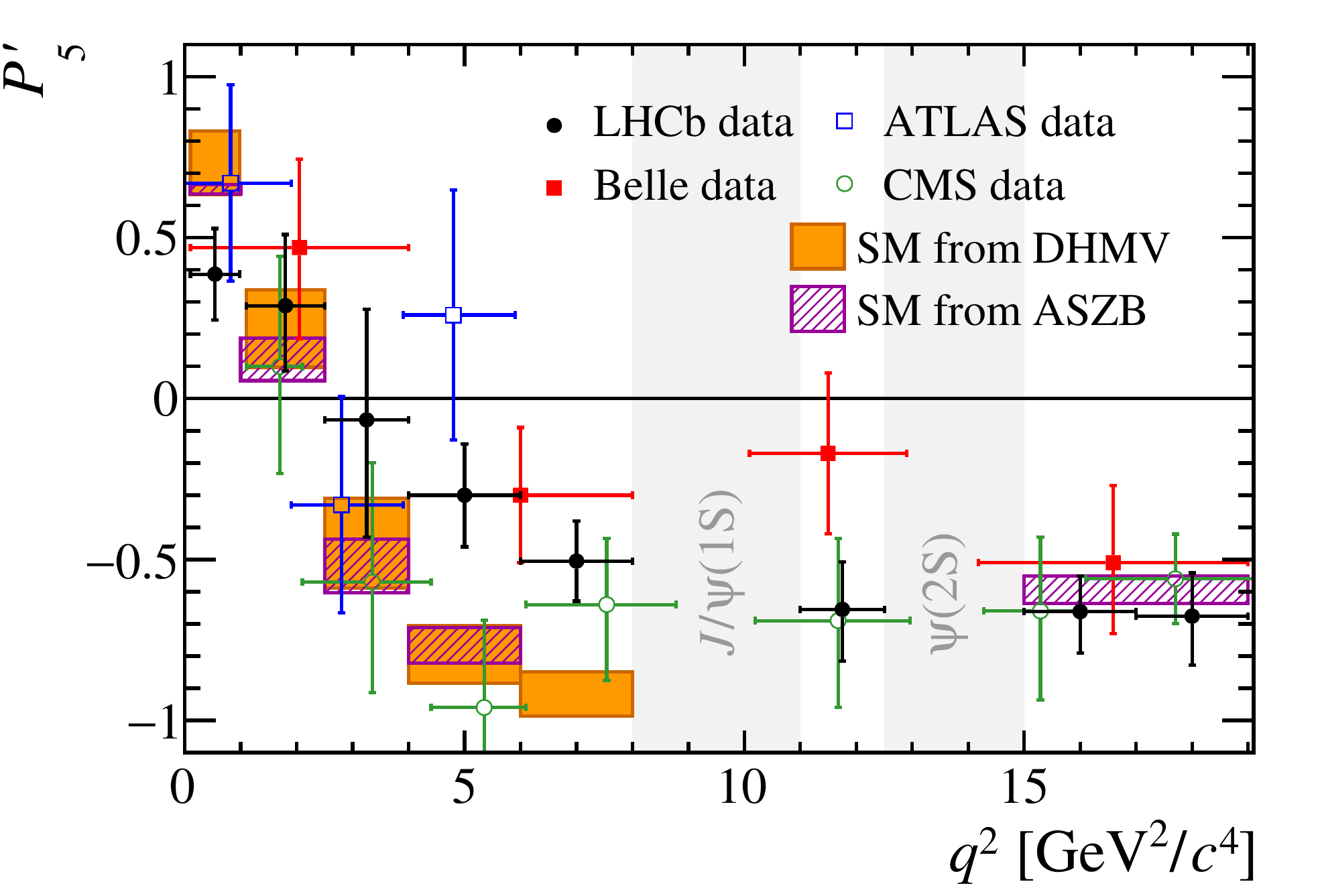}}
\raisebox{-.5\height}{\includegraphics[width=.4\textwidth]{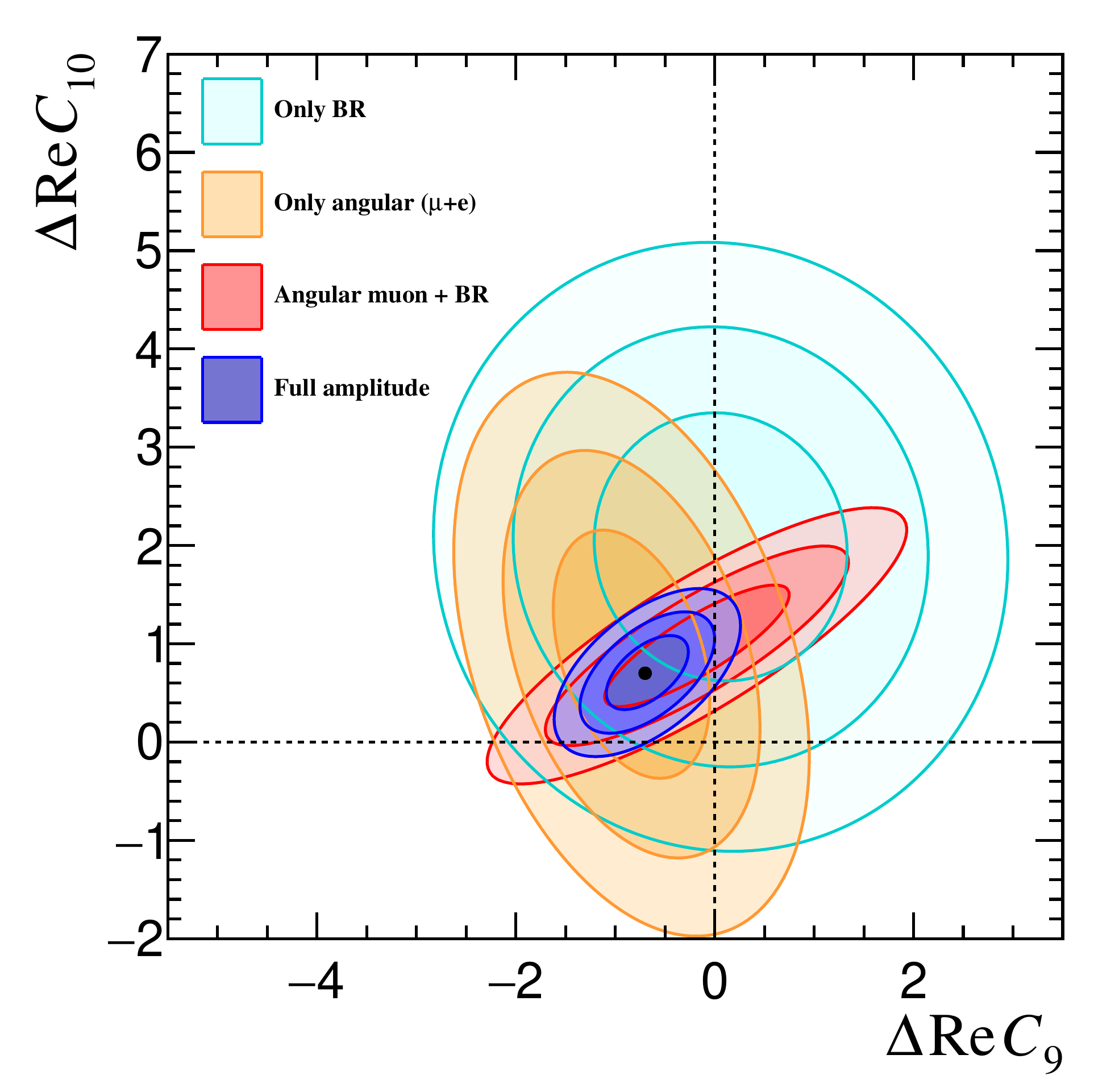}}
\caption{The $P_5'$ observable measured by the LHCb~\cite{Aaij:2015oid}, CMS~\cite{Sirunyan:2017dhj}, Belle~\cite{Wehle:2016yoi} and ATLAS~\cite{Aaboud:2018krd} experiments, superimposed to the SM predictions (left). The Belle result includes both $\mu\mu$ and $ee$ data. The largest tension is registered in the dimuon channel~\cite{Wehle:2016yoi}. Expected sensitivity to NP contributions in $\mathcal{C}_9$ and $\mathcal{C}_{10}$, shown as $1$, $2$ and $3\sigma$ countours, after the LHC Run 2 (right)~\cite{Mauri:2018vbg}.}\label{fig:p5p}
\end{figure}

The CP-averaged differential $B^0\to K^{*0}(\to K^+\pi^-)\mu^+\mu^-$ decay rate in terms of the three angular observables $\cos\theta_K$, $\cos\theta_\ell$, $\phi$ and of the dilepton invariant mass $q^2$ can be written as
\begin{align}
\frac{d^4\left(\Gamma + \bar\Gamma\right)}{dq^2d\vv\Omega} = \sum_i I_i(q^2) f_i(\vv\Omega)
\end{align}
where $\Gamma$ and $\bar\Gamma$ denote decays of a $b$ and $\bar b$ quark, respectively, $f_i$ are combinations of spherical harmonics and the $I_i$ are $q^2$-dependent angular observables.
The latters comprise CP-even ($S_i$) and CP-odd ($A_i$) observables, and are sensitive to the $\mathcal{C}_9$ and $\mathcal{C}_{10}$ Wilson coefficients. A total of 15 observables arise, due to the interplay between the vector ($K^{*0}$) and scalar contributions to the $K\pi$ system.
From these observables, optimized quantities $P_i^{(\prime)}$ can be constructed, for which the $B^0\to K^{*0}$ form factor uncertainties cancel at leading order.
LHCb measured these angular observables using an unbinned maximum likelihood fit to the three angular spectra, $m(K\pi)$ and $m(K\pi\mu\mu)$, and applying an independent method as a cross-check~\cite{Aaij:2015oid}. This measurement is performed in $q^2$ ranges chosen in order to exclude charmonium resonances.
The reconstructed $B^0$ mass is used to discriminate between signal and background, whereas the $K\pi$ mass distribution is exploited to separate vector and $S$-wave contributions.
\begin{figure}
\centering
\includegraphics[width=.325\textwidth]{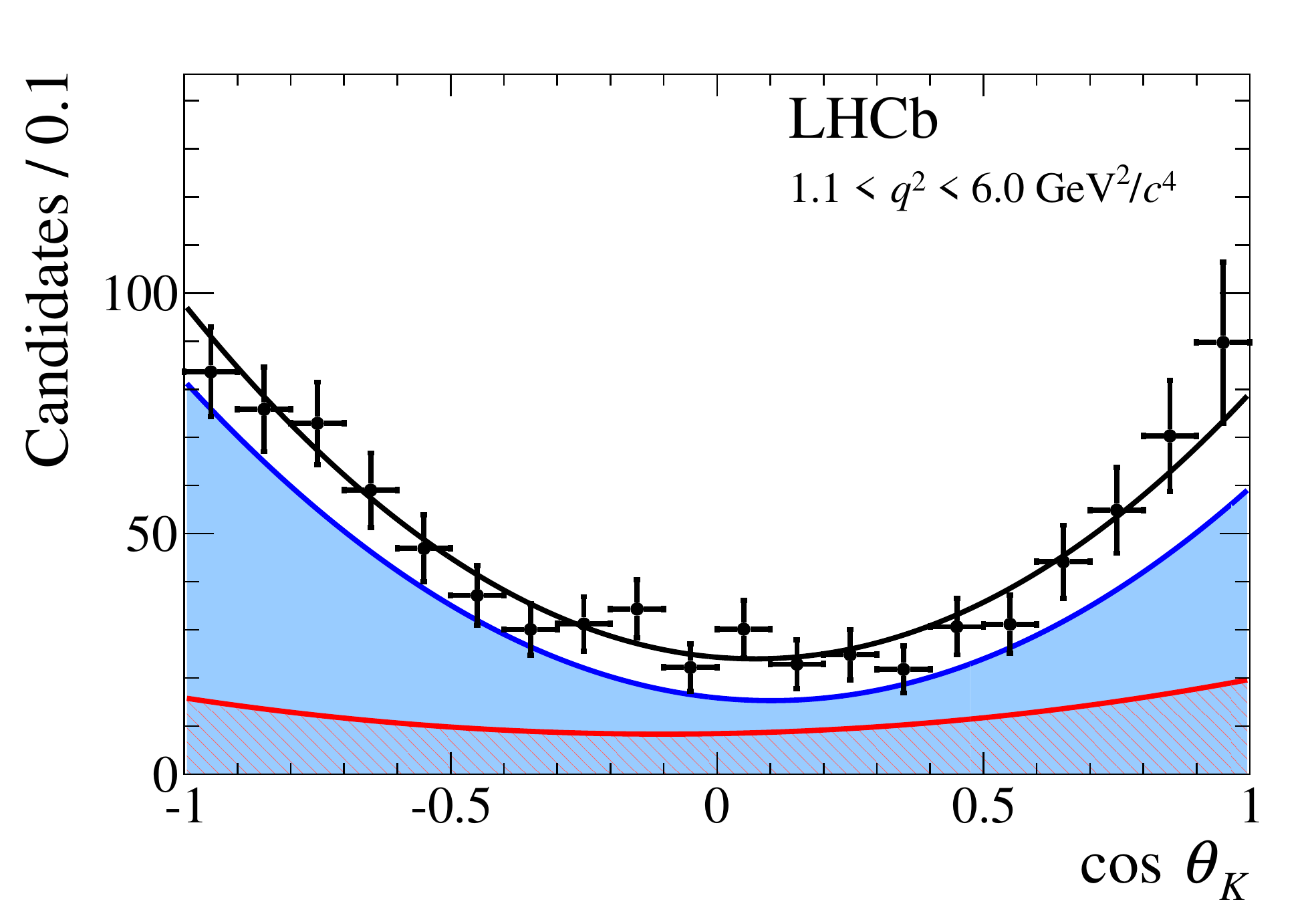}
\includegraphics[width=.325\textwidth]{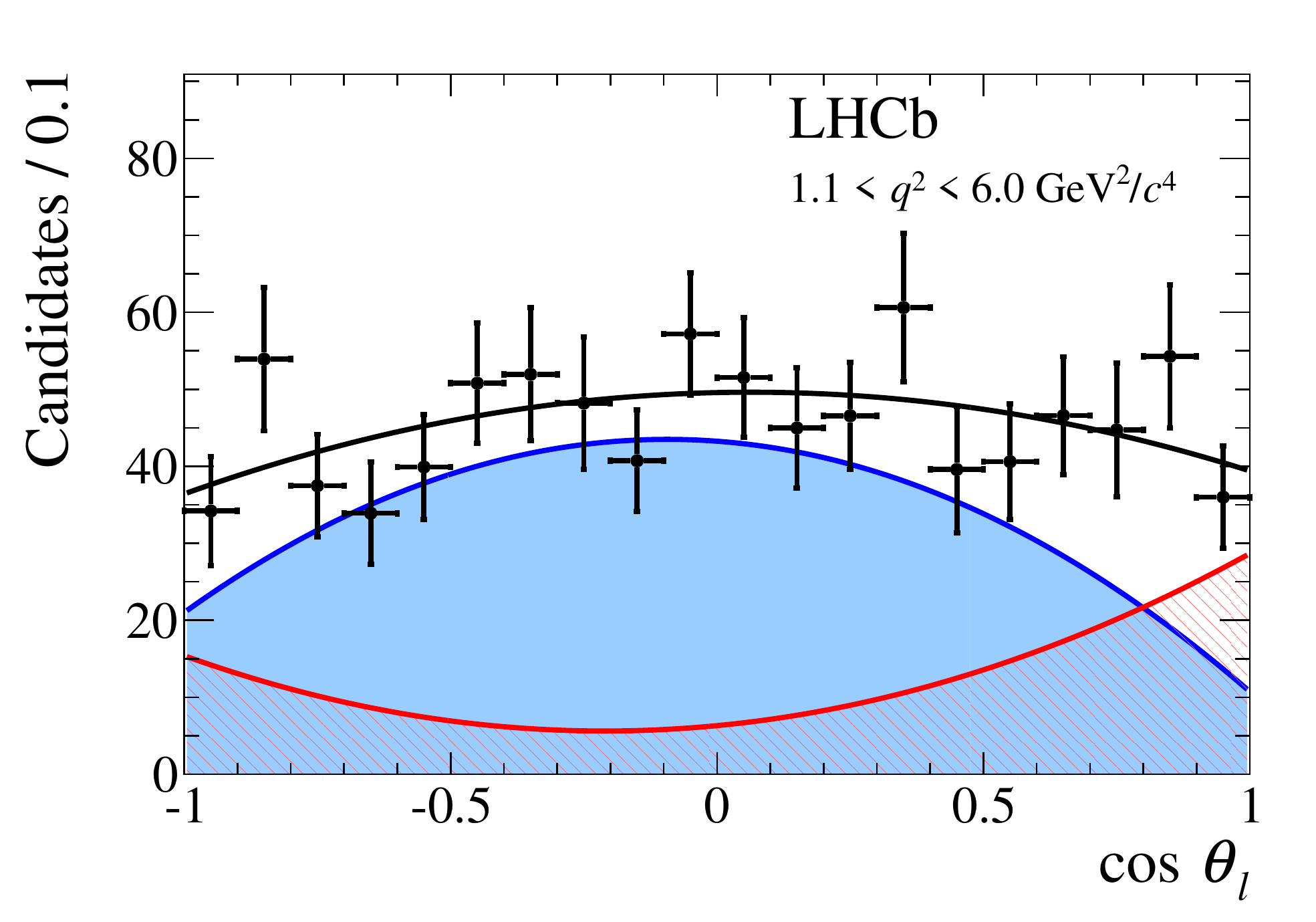}
\includegraphics[width=.325\textwidth]{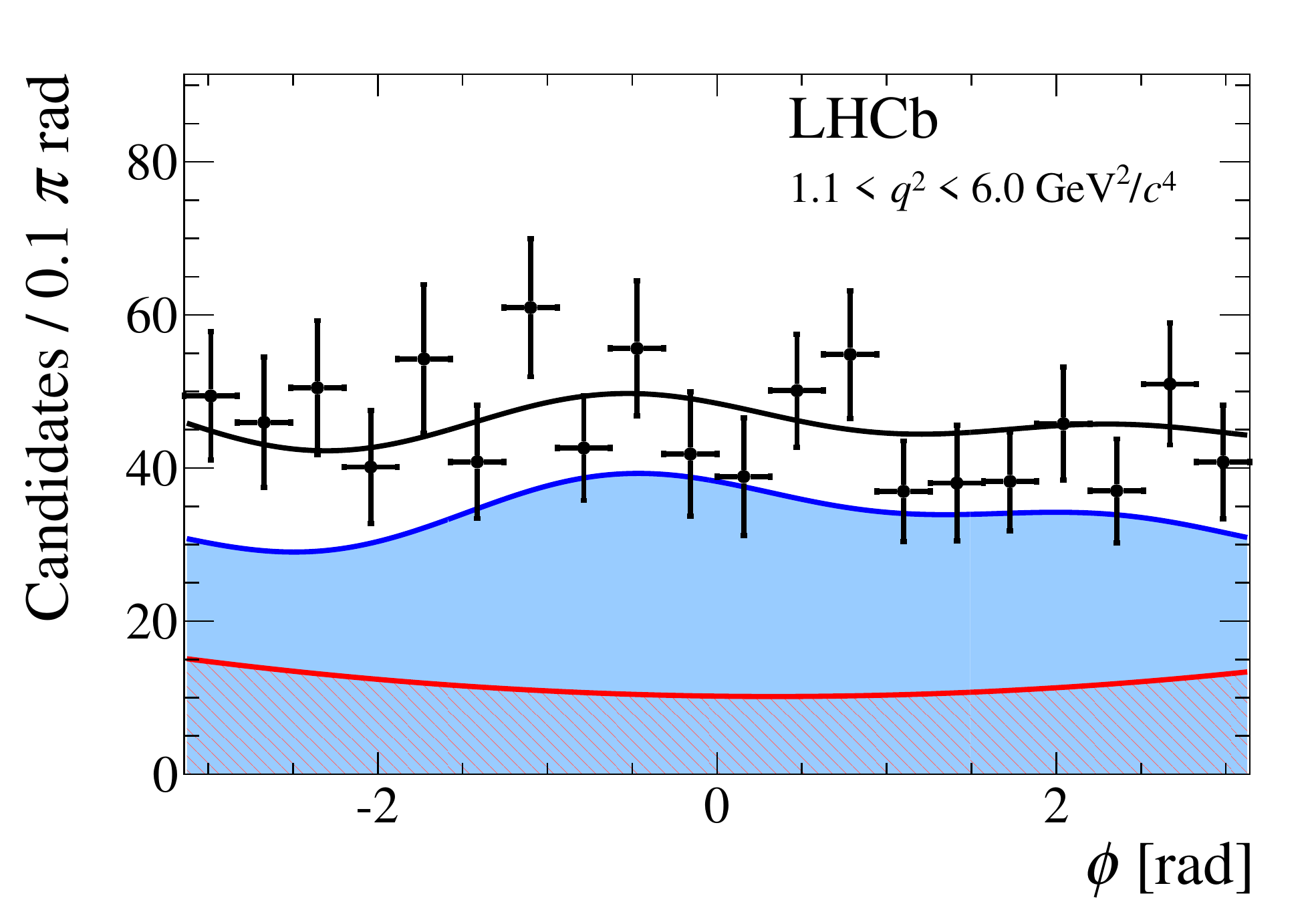}
\caption{Fit of LHCb data to the three angles describing the $B^0\to K^{*0}(\to K\pi)\mu\mu$ decay
in a central $q^2$ bin, where signal and background contributions are depicted in blue and red~\cite{Aaij:2015oid}.}\label{fig:kstmumufits}
\end{figure}
The fit results for a central $q^2$ bin are shown in \figurename~\ref{fig:kstmumufits}.
Most observables are in good agreement with the SM. However, a significant tension ($3.4\sigma$) is observed in the optimised observable $P_5'$.
This tension has been registered also by the Belle~\cite{Wehle:2016yoi} and ATLAS~\cite{Aaboud:2018krd} experiments, although with larger uncertainties. A CMS analysis~\cite{Sirunyan:2017dhj} results compatible with both the LHCb result and the SM. An overview of the $P_5'$ measurements is shown in \figurename~\ref{fig:p5p} (left).

\section{Interpretation and prospects}
Global fits to LFU tests in $b\to s\ell\ell$ decays highlight $4\sigma$ tensions in $\mathcal{C}_9$ and $\mathcal{C}_{10}$, when NP is constrained to contribute to a single Wilson coefficient~\cite{Altmannshofer:2017yso}.
These tensions reach a level of $5\sigma$ if other observables, such as the angular coefficients and branching fractions of $b\to s\mu\mu$ decays are included in the global fit~\cite{Capdevila:2017bsm}; however these observables have much larger theoretical uncertainties.
It has been suggested that an incorrect evaluation of long-distance effects from vector charmonium contributions could be the responsible for some of the observed discrepancies. However, an LHCb measurement of the interference between long- and short-distance effects in $B^+\to K^+\mu\mu$ suggests that such and effect may not be sufficient to explain the observations~\cite{Aaij:2016cbx}.

A coherent picture emerges from the tensions observed in $b\to c\ell\nu$ and $b\to s\ell\ell$ transitions.
Both sets of anomalies have a significance in the range of $4\sigma$.
The large difference between tree-level and loop-level amplitudes, the significance and weight of the anomalies, and the fact that no deviations from theory have been observed so far in decays of light mesons prompted the physics community to develop NP models with particles that
couple preferentially to the second and third generation, in a Yukawa-like hierarchy~\cite{Buttazzo:2017ixm,Wei:2017ago,Bauer:2015knc,Greljo:2015mma,Barbieri:2017tuq,Bordone:2017bld}.
Direct searches for such new mediators have been performed at CMS~\cite{Sirunyan:2018jdk,CMS-PAS-EXO-18-008} and ATLAS~\cite{Aaboud:2016qeg}, so far without success. Searches for lepton flavour violating decays, also predicted by such models, are reaching unprecedented sensitivities~\cite{Aaij:2015qmj,Aaij:2014azz,Aaij:2017cza}.

A recent work~\cite{Mauri:2018vbg} found that a simultaneous analysis of $B^0\to K^{*0}\mu^+\mu^-$ and $B^0\to K^{*0}e^+e^-$ amplitudes has the potential of turning the anomalies into a groundbreaking discovery already with the LHC Run 2 dataset, as shown in the right-hand panel of \figurename~\ref{fig:p5p}. Measurements from the newly started Belle 2 run are also expected to shed light on the current anomalies, with the added reliability of a complementary experimental setup. For example, the LHCb uncertainty on the \RDst{} ratio is expected to scale down about a factor 0.5 with the LHC Run 3, and Belle 2 will have enough data by then to provide an \RD{} measurement with an uncertainty 2 to 3 times smaller than the current world average~\cite{Albrecht:2017odf}. If the flavour anomalies persist, striking evidence of new physics will be available on a short time scale.

\section*{References}
\bibliography{bib}

\end{document}